\documentclass[aps,prd,
nofootinbib,
reprint,
groupedaddress,amsmath,amssymb,showpacs]{revtex4-1}

\usepackage{graphicx}
\usepackage{bm}
\usepackage{dcolumn}
\def \pbar{\bar{p}}

\begin{document}

\title{Classical color field modified minijet model for $pp$ and $\pbar p$ total cross section}


\author{\textsc{Man-Fung Cheung}}
\email[]{mfcheung@physics.utexas.edu}
\author{\textsc{Charles B. Chiu}}
\email[]{chiu@physics.utexas.edu}
\affiliation{Center for Particles and Fields and Department of Physics\\
University of Texas at Austin, Austin, TX 78712, USA}


\date{\today}

\begin{abstract}
In a recent paper, we have evaluated the $gg\rightarrow gg$ scattering amplitude in the presence of classical color field generated by the colliding protons in the leading order approximation within the pQCD.  In this work, we show that this amplitude can be resumed to obtain the classical color field modified $gg \rightarrow gg$ elastic scattering amplitude.  This modified amplitude is suppressed when the longitudinal momentum fraction, $x$, of the incident gluon is small.  Minijet cross section is calculated using the modified amplitude.  We show that the $pp$ and $\pbar p$ cross section from $\sqrt{s} = 5$ GeV to $30$ TeV can be described as a sum of a hard component contributed by the modified minijet model and a soft component due to the exchange of the pomeron and of the $I=0$ exchange-degenerate $\omega$ and $f$ trajectories.  The predicted cross section has a $\ln^2 s$ asymptotic behavior which satisfies the Froissart bound.  
\end{abstract}

\pacs{13.85.-t, 13.75.Cs, 13.60.Hb}

\maketitle


\section{\label{sec:1}Introduction}
It is a well accepted notion that QCD is the underlying theory of hadron physics.  However, computing total cross sections such as $pp$ and $\bar{p} p$ cross sections in the large CM collision energy, $\sqrt{s}$, remains an unresolved problem in QCD.  Unlike hard processes which can be computed in the perturbation theory, to compute the total cross section requires knowledge of the imaginary part of the forward elastic scattering amplitude, which involves the intrinsically nonperburative zero momentum transfer physics.  In the 60's and 70's Regge theory was extensively developed in attempts to understand hadron interactions \cite{Chiu1972,Fox1973,Collins1977}\footnote{For application of Regge theory in late 60's and the early 70's see the two review articles \cite{Chiu1972,Fox1973} and the book \cite{Collins1977}.  For more recent discussion on the Regge theory and its relation to QCD see the book \cite{Donnachie2002}.  See also those references listed therein.}.  Even now one still cannot claim to have an understanding of the total cross sections from first principles of QCD \cite{Donnachie2002}.  

We recall the minijet model was first introduced in the 70's \cite{Cline1973}.  At the time it was noticed that the rise of the total cross section was very similar to the jet production cross section. In this context, it is natural to separate the total cross section into a soft component and a hard component. Here the hard component is to be computed through the pQCD motivated minijet model.  However, the minijet cross section rises too rapidly with respect to $\sqrt{s}$.  Since then the minijet model has been incorporated in the eikonal model by various authors in attempts to tame the rise and explain the data quantitatively \cite{Pancheri1984,Gaisser1985,Durand1987, Capella1987,Luna2005,Godbole2005,Achilli2011}.  

In this paper, we compute of minijet cross section using pQCD and demonstrate the taming of the rapid rise within the framework of QCD using the classical color field modified amplitude.  
We find the modified minijet model can qualitatively describe the total cross section over the entire range of the available data, i.e. from $\sqrt{s}=5$ GeV to 30 TeV.  

The remainder of the paper is organized as follows.  In Sec. \ref{sec:2}, we define the conventional minijet cross section and discuss its violation of the Froissart bound (FB).  Then we present the classical color field modified minijet model in Sec. \ref{sec:3}.  In Sec. \ref{sec:4}, we compare the present model with the data.  Asymptotic behavior of the present model will also be discussed.  This work is concluded in Sec. \ref{sec:5}.  Details on the derivation of the modified $gg\rightarrow gg$ amplitude used in the present model can be found in \cite{Cheung2011a}.  

\section{\label{sec:2}Minijet model and violation of the Froissart bound}
Following the soft and hard decomposition of the total cross section we write $pp$ and $\bar{p} p$ cross sections as  (see also \cite{Godbole2005,Achilli2011})
\begin{eqnarray}
\sigma_{pp}&&= \sigma_{soft}+\sigma_{hard} \label{eq:softpp} \\ 
\sigma_{\bar{p} p}&&= \sigma_{soft}(1+a/\sqrt{s}) + \sigma_{hard}  \label{eq:softpbp}
\end{eqnarray}
For the soft component, we assume Regge theory is applicable.  Here the dominant contribution is the Pomeron exchange with $\alpha_P(0)=1$.  So $\sigma_{soft}=\mbox{const} \times s^{\alpha_P(0)-1} =\mbox{const}$.  The secondary Regge pole contribution is expected to be dominated by $I=0$ exchange of the $f$ and $\omega$ trajectories.  We assume they are exchange degenerate trajectories \cite{Arnold1965,Henzi1967,Harari1968,Chiu1968}\footnote{The idea of exchange degeneracy was first introduced in \cite{Arnold1965}.  Soon after that, it was recognized the exchange of pairs of exchange-degenerate trajectories play a crucial rule in hadrons collision phenomenology in intermediate energy region, especially in the context of direct-channel and crossed-channel duality.  For instance the presence of prominent low energy meson resonances in the $\bar{p} p$ channel is responsible for the presence of the $1/\sqrt{s}$ term in eq. (\ref{eq:softpbp}).  For the $pp$ case, the lack of low energy resonances leads to the absence of $1/\sqrt{s}$ term in eq. (\ref{eq:softpp}).  For discussions on exchange degeneracy and its relation duality see \cite{Henzi1967,Harari1968,Chiu1968}.} with $\alpha_f (0)=1/2$ and $\alpha_\omega(0)=1/2$.  (Note that here the pomeron is associated with the soft component contribution with $\alpha_P(0) = 1$.  This differs from the pomeron used in \cite{Donnachie1992}.  The latter is associated with both soft and hard contributions with an intercept at $\alpha_P(0)=1.08$.) 

The hard component has been associated with the presence of jets in the final state, where the inclusive jet cross section is assumed to be dominated by the minijet production contribution \cite{Cline1973,Pancheri1984,Gaisser1985}.  At high energy, the dominant minijet cross section comes from the $gg\rightarrow gg$ process and is given by:
\begin{eqnarray}
\sigma_{mnj} = \int_{\frac{2\hat{t}_0}{s}}^{1} && dx_1 \int_{\frac{2\hat{t}_0}{x_1s}}^{1}dx_2 \int_{-\hat{s}+\hat{t}_0}^{-\hat{t}_0}d\hat{t}\, \nonumber \\ && \times g(x_1,\mu_g^2)\, g(x_2,\mu_g^2) F_{mnj}\frac{d\sigma_{gg}}{d \hat{t}}, \label{eq:mnj}
\end{eqnarray}
where $x_1$ and $x_2$ are the longitudinal momentum fractions of the gluons from the two colliding protons.  $F_{mnj}$ is introduced as a generalization factor which will be elaborated in later sections.  For the conventional minijet model, $F_{mnj} = 1$.  The differential cross section of the gluon-gluon elastic scattering at tree level is given by $\frac{d\sigma_{gg}}{d\hat{t}} = \frac{\pi\alpha_s^2(\mu_g^2)}{\hat{s}^2} |M|^2$, where
\begin{equation}
|M|^2 = \frac{9}{2}\left(3-\frac{\hat{u}\hat{t}}{\hat{s}^2} -\frac{\hat{s}\hat{t}}{\hat{u}^2} -\frac{\hat{u}\hat{s}}{\hat{t}^2} \right) \approx \frac{9}{2}\left(-\frac{\hat{s}\hat{t}}{\hat{u}^2} -\frac{\hat{u}\hat{s}}{\hat{t}^2} \right).\label{eq:sigmagg}
\end{equation}
The quantities $\hat{s}$, $\hat{t}$ and $\hat{u}$ are the Mandelstam variables of the subprocess that $\hat{s}=x_1x_2s$, $\hat{t} = q^2$ and $\hat{u}=-\hat{s}-\hat{t}$, where $q$ is the momentum exchange in the subprocess.  The gluon distribution function of the incident protons $g(x,\mu_g^2)$ are evaluated at the scale $\mu^2_g= \hat{t}\hat{u}/\hat{s} = p_T^2$, where $p_T$ is the transverse momentum of the gluons in the final state.  The parameter $\hat{t}_0$ is the cutoff of the squared exchange momentum which defines the hard scattering scale below which reactions are considered to be soft.  One can verify that for large $s$ the dominant contributions of the $gg$ subprocess are from the terms with $\hat{t}$ and $\hat{u}$ singularities.  The corresponding leading order diagrams are the one gluon exchange in $\hat{t}$- and $\hat{u}$-channel which lead to the approximation in eq. (\ref{eq:sigmagg}).

\subparagraph*{Violation of the Froissart bound:}
To examine the asymptotic behavior of $\sigma_{mnj}$, one can approximate $\sigma_{mnj}$ by its most dominant contribution.  Firstly, by peak approximate of the $\hat{t}$ and $\hat{u}$ singularities, we replace all the $\mu^2_g$ by $\hat{t}_0$ and $\frac{d\sigma_{gg}}{d\hat{t}}\rightarrow 9\pi \alpha_s(\hat{t}_0)/\hat{t}^2$.  Thus, the integral of $\hat{t}$ is asymptotically $s$-independent.  The integrand of eq. (\ref{eq:mnj}) in the small-$x$ region dominates since $g(x,\mu^2_g)$ increases rapidly as $x$ decreases.  A conventional power law (PL) parametrization of $g(x,\mu^2_g)$ is $g\sim x^{-J}$ when $x$ is small.  The asymptotic form of $\sigma_{mnj}$ becomes
\begin{eqnarray}
\sigma^{PL}_{mnj}(s) 
&\xrightarrow{s \rightarrow \infty} \int_{\frac{2\hat{t}_0}{s}}^1 dx_1 \int_{\frac{2\hat{t}_0}{x_1 s}}^1 dx_2 \frac{1}{(x_1 x_2)^{J}} \propto s^{J-1}\ln s .\label{eq.sigmajet2}
\end{eqnarray}
This behavior is in agreement with that given in an earlier work (See eq. (18) of \cite{Godbole2005}).  The deep inelastic scattering data suggests $J\sim 1.16$ to $1.42$ \cite{Chekanov2010} so that 
at large $s$, $\sigma^{PL}_{mnj}$ violates the Froissart bound \cite{Froissart1961,Martin1963}, which requires that $\sigma\le \mbox{const.} \times \ln^2 s$.

Recently, there is a alternative parametrization for the gluon distribution derived directly from a Froissart bound satisfying fit of the proton structure function $F_2(x,Q^2)$ \cite{Block2008}.  The distribution is a quadratic polynomial  in $\ln (1/x)$ with quadratic polynomial coefficients in $\ln Q^2$.  Explicitly, for $0<x<0.09$, it reads  
\begin{eqnarray}
&&xg^{FB}(x,Q^2) =  -0.459 - 0.143 \ln Q^2 -0.0155 \ln ^2 Q^2 \nonumber \\
&& + [0.231 + 0.00971 \ln Q^2 - 0.0147 \ln ^2 Q^2] \ln(1/x)\nonumber \\
&&+[0.0836 + 0.06328 \ln Q^2 +0.0112 \ln ^2 Q^2]\ln^2(1/x).
\label{eq:gFB}
\end{eqnarray}
The asymptotic behavior of the minijet calculated using $g^{FB}(x,Q^2) \sim (1/x)\ln^2 (1/x)$ is 
\begin{eqnarray}
\sigma^{FB}_{mnj} \approx \int_{\frac{2\hat{t}}{s}}^1dx_1 \int_{\frac{2\hat{t}}{x_1s}}^1dx_2 \frac{\ln^2(x_1) \ln^2(x_2)}{x_1 x_2} \propto \ln^6 s
\end{eqnarray}
Despite the corresponding $F_2$ satisfying Froissart bound, the minijet cross section calculated with $g^{FB}(x,Q^2)$ does not.  Similar violation has been shown in neutrino-nucleon deep-inelastic scattering \cite{Illarionov2011}.  In fact, the authors in \cite{Block2008} pointed out that one should not expect a leading order approximation to be constrained by Froissart bound \cite{Block2011b}.  

\section{\label{sec:3} Minijet in classical color field}
In the original minijet model, each incoming proton provides a gluon undergoing elastic scattering in vacuum to produce the jets.  For each subprocess, the rest of the protons are treated as spectators.  In contrast to this simple picture, we consider the effect due to the other gluons in the protons.  In particular, the large $x$ and small $x$ gluons inside the protons are treated collectively as a classical source and a classical color field, respectively, as in the color glass condensate (CGC) (for a recent review, see \cite{Gelis2010}).  However, in this work, the classical field, instead of being a degree of freedom in the problem, is considered as a prescribed background of which properties are characterized by the rapidity of the collinear gluons in the scattering.  For a $gg$ subprocess where the incident gluons have longitudinal momentum fraction $x_1$ and $x_2$, $x_1$ and $x_2$ define the separation scales between the fast and slow gluons in the protons.  Since $x_1$ and $x_2$ are independent variables and, to the leading order, the solution of the classical field is the superposition of the individual field generated by each proton.  The effect of the total classical field on the amplitude is expected to be a product of the individual effect.  First consider the gluon with $x_1$ from the proton moving along the $+z$ direction.  The gluons in the proton with $x>x_1$ are regarded as fast and treated as random classical color sources, $\rho_1$, moving along the light-cones.  The gluons with $x<x_1$ are treated as a classical color field, $A$, generated by the source through the Yang-Mills equation.  The field $A$ can be solved in terms of $\rho_1$.  (The same situation appears in the other proton.)  Therefore, the gluons are scattering in a background with classical color field instead of the vacuum.  

The classical color field interacts with the quantum gluons, $B$, involved in the hard scattering process and modifies the $gg\rightarrow gg$ amplitude.  Due to the color neutrality assumption, the overall classical color field must be zero, but the fluctuation can be finite.  So the leading order contribution of the classical field to the amplitude is proportional to $A^2$; therefore, $B$ interacts at least twice with $A$.  The modified amplitude depends on $A[\rho_1]$, in turn, the random sources $\rho_1$.  To obtain a physical amplitude, the amplitude must be averaged over the source with a weight function $W[\rho_1]$.  For that purpose, we apply the Gaussian average of the IIM model \cite{Iancu2002b}.  Although the Gaussian average is an approximation to the full renormalization group solution of the weight function, it captures the physics in both dilute and saturated regime.  

We refer the reader to \cite{Cheung2011a} for the detailed derivation of the modified $gg\rightarrow gg$ amplitude.  Here we quote the result of the modified amplitude.  In particular, for the $\hat{t}$ and $\hat{u}$-channel one gluon exchange diagrams at high energy, the propagator of the exchanged quantum gluon is modified while the vertex remained unchanged.  The modification to the propagator due to the proton moving to the $+z$ direction to the leading order in the coupling and the classical field contains two terms.  Under the eikonal approximation at high energy scattering that the momentum exchange $q^\mu \ll \sqrt{\hat{s}}$, only one of the modification terms in \cite{Cheung2011a} is relevant.  The modified propagator is given by 
\begin{eqnarray}
G_{\mu\nu}^{ab}(q,x_1) = &&g_{\mu\nu}\delta^{ab} \frac{-i}{q^2} \nonumber \\
&& -g_{\mu\nu}\delta^{ab}\frac{\alpha_s N_c}{\pi L^-} I(Q^2,x_1)sgn(q^-)\frac{q^-}{q^4}, \label{eq:G}.
\end{eqnarray}
The function $sgn(x) = \theta(x)-\theta(-x)$ and 
\begin{equation}
I(Q^2,x_1) = \int_{\frac{1}{R_p^2}}^{Q^2}\frac{d^2k_\perp}{k_\perp^4} \mu(k_\perp^2,x_1),
\end{equation}  
where $\mu(k_\perp^2,x_1)$ is given by the IIM model as
\begin{equation}
\mu(k_\perp^2,x)= \frac{k^2_\perp}{\gamma c \alpha_s N_c} \ln \left[1+ \left(\frac{Q_s^2(x)}{k_\perp^2}\right)^\gamma\right],
\end{equation}
$Q^2= -q^2$ and $R_p$ is the radius of proton which is set to be $R_p = 0.8768$ fm.  The saturation scale, $Q_s^2(x)$, characterized the density of the source and has a power law dependence on $x$.  For $x\sim 1$, the source is dilute and $Q_s^2(x)$ is expected to vanish.  In this  analysis, we assume 
\begin{equation}
Q_s^2(x) = Q_0^2 \left(\frac{x_0}{x}\right)^{\lambda} (1-x)^p, 
\end{equation}
where the value of $Q_0^2$, $x_0$ and $\lambda$ are set to be the typical values from the analysis of DIS \cite{Golec-Biernat1998, Iancu2004} as $Q_0^2 = 1$ GeV$^2$, $x_0=3\times 10^{-4}$ and $\lambda = 0.28$.
We use $p=5$ in the calculation since in the dilute region, $Q_s^2 \propto xg(x,Q^2) \approx (1-x)^p$ and the exponent, $p=5$, is motivated by the simple choice of $xg$ in \cite{Bartels2002}.  In fact, the value of $p$ is insensitive to the cross section since there is a very small contribution from the region where $x\sim 1$.  The function $I(Q^2,x)$ can be integrated analytically as
\begin{eqnarray}
I(Q^2,x) = \frac{\pi}{\alpha_s N_c \gamma c} &&\left[ Li_2\left(-\left(\frac{Q_s^2(x)}{Q^2}\right)^\gamma\right) \right. \nonumber \\ 
&&\left. -Li_2\left(-\left(Q_s^2(x) R_p^2\right)^\gamma\right) \right],\label{eq:I}
\end{eqnarray}
where $Li_2(x)$ is the dilogarithm.  $I(Q^2,x)$ is a monotonically increasing function in $Q_s^2(x)$ and $I(Q^2,x=1)=0$.  

\begin{figure}[h]
\centering
\includegraphics[height=.8in]{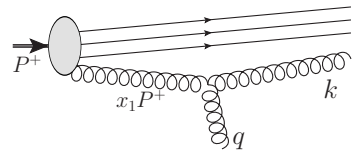}
\caption{Momentum conservation of the vertex and on-shell condition for the incoming($x_1P^+$) and outgoing($k$) gluons: $-2x_1 P\cdot q + q^2=0 \Rightarrow q^- = q^2/(2x_1 P^+)$.}
\label{fig:1}
\end{figure}

Eq. (\ref{eq:G}) can be further simplified.  The parameter $L^-$ is the average longitudinal size of the source with x for $x_1<x<1$.  We defined $L^- = \frac{\chi}{x_1 P^+}$.  One can roughly estimate a reasonable value of $\chi$ for small-$x$ and low $Q^2$ with $g(x,Q^2)$.  For example, using $g\sim x^{-J}$ in \cite{Chekanov2010} with $J=1.2$ at $Q^2=1.8$ GeV$^2$, 
\begin{eqnarray}
L^- \equiv \langle \frac{1}{xP^+} \rangle  = \frac{\int_{x_1}^1 \frac{x^{-J}}{xP^+} dx}{ \int_{x_1}^1 x^{-J} dx} \approx \frac{J-1}{J}\frac{1}{x_1P^+} \label{eq:L}
\end{eqnarray}
Therefore, $\chi = \frac{J-1}{J}\approx 0.17$.  $\chi=0.168$ is used in our calculation.  Moreover, when one applies the propagator to the $\hat{t}$ and $\hat{u}$ channel exchanges, the energy-momentum conservation of the vertex and the on-shell condition of the incoming and outgoing gluons requires $q^- = q^2/(2x_1P^+)$ as we show in Fig.\ref{fig:1}.  
$q^-$ is negative as in physical region $q^2<0$.  Using eq. (\ref{eq:L}), $q^- = q^2 L^- /(2\chi)$.  The $L^-$ dependence in the denominator of eq. (\ref{eq:G}) and in $q^-$ are canceled.  The propagator becomes
\begin{equation}
G_{\mu\nu}^{ab}(q,x_1) = g_{\mu\nu}\delta^{ab}\frac{-i}{q^2} \left[ 1  + i\frac{\alpha_s N_c}{2 \chi \pi} I(Q^2,x_1)\right] \label{eq:G2}
\end{equation}
and has simple form of a Feynman propagator multiplied by a pure imaginary factor that depends on $x_1$ and $Q^2$.  

\begin{figure}[h]
\centering
\includegraphics[height=0.45in]{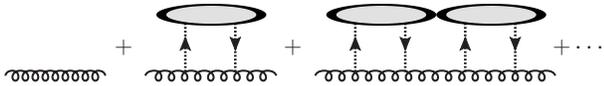}
\caption{\label{fig:2}Schematic Feynman diagram which represents the iterative sum of the classical field modified propagator.  The first term is the bare propagator.  The blob and the two lines connecting to it in the series represent the interaction between the quantum gluon propagator and the classical field.}
\end{figure}

\subparagraph*{\label{sec:3a} Iterative sum and classical field modification factor:}
The final modified propagator is given by the iteration of the correction in eq. (\ref{eq:G2}).  Diagrammatically, it corresponds to the series in Fig. \ref{fig:2}.  Each blob represents the gluon's interaction with the $A$ field from the proton and contributes the same multiplicative factor.  
The sum of the series is a geometric sum.  So the final form of the propagator is the product of the bare propagator and a correction factor,
\begin{eqnarray}
G_{\mu\nu}^{ab}(q,x_1) = &&g_{\mu\nu}\delta^{ab}\frac{-i}{q^2} \left[ 1  + i\frac{\alpha_s N_c}{2 \chi \pi}I(Q^2,x_1) \right. \nonumber \\
&&+\left. \left( i\frac{\alpha_s N_c}{2 \chi \pi} I(Q^2,x_1)\right)^2 + \cdots \right]\nonumber \\
=&& g_{\mu\nu}\delta^{ab}\frac{-i}{q^2}\left[ \frac{ 1}{1-  i\frac{\alpha_s N_c}{2 \chi \pi}I(Q^2,x_1)} \right]\label{eq:Gfull}
\end{eqnarray}
The expression of eq. (\ref{eq:Gfull}) is analytic on the complex plane of $I(Q^2,x_1)$ except one point.  According to analytic continuation, one can analytically continues the propagator to a region where $I(Q^2,x_1)$ is larger, therefore, equivalently, the large $Q_s^2(x_1)^2$ or small $x_1$ region.

\begin{figure}[h]
\centering
\includegraphics[width=0.7\linewidth]{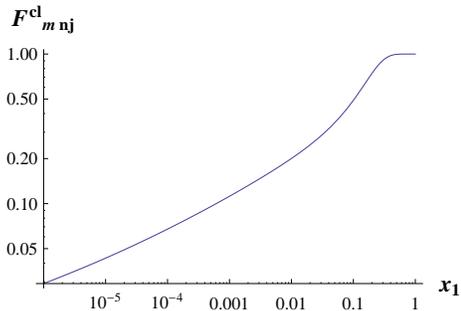}
\caption{\label{fig:3} $x_1$-dependence of $F^{cl}_{mnj}(x_1,x_2,Q^2)$ for $x_2=1$ and $Q^2=1$ GeV$^2$.}
\end{figure}

For the case of the other proton providing the gluon with $x_2$, a similar correction factor is introduced to the propagator, except that the indexes  $-$ and $1$ are changed to $+$ and $2$; namely, $\{L^-,q^-,x_1\} \rightarrow \{L^+,q^+,x_2\}$.  Going through the same analysis for $L^-$ and $q^-$, the modified propagator is obtained as
\begin{eqnarray}
G_{\mu\nu}^{ab}(q,x_1) =  g_{\mu\nu}\delta^{ab}\frac{-i}{q^2}f(Q^2,x_1)f(Q^2,x_2), \label{eq:Gfull2}
\end{eqnarray}
where 
\begin{equation}
f(Q^2,x) = \frac{ 1}{1-  i\frac{\alpha_s N_c}{2 \chi \pi}I(Q^2,x)}.
\end{equation}
Applying to the minijet model, the correction factor is multiplied to the $\hat{t}$ and $\hat{u}$ channel amplitudes as a numerical factor.  It is equivalent to multiplying a factor $F^{cl}_{mnj}(x_1,x_2)$ to the $gg \rightarrow gg$ differential cross section, where   
\begin{equation}
F_{mnj}^{cl} \equiv |f(Q^2,x_1)f(Q^2,x_2)|^2 \label{eq:Fcl}.
\end{equation} 
We refer to $F_{mnj}^{cl}$ as the \textit{classical color field modification factor}.  The $x_1$-dependence of $F_{mnj}^{cl}$ is shown in Fig. \ref{fig:3}.  If both $x_1$ and $x_2$ equal to one, $F_{mnj}^{cl} = 1$ and the conventional minijet is recovered.  When $x_1$ or $x_2$ becomes small, $F_{mnj}^{cl}$ is less than $1$.  Therefore, the contribution from the high gluon density region, where $x_1$ or $x_2$ is small, is suppressed.  

\section{\label{sec:4}Implication for $pp$ and $\bar{p} p$ total cross section at high energy}
\subparagraph*{\label{sec:4a}Comparing to data:}
To compute the minijet cross section at high energy, we approximate $d\sigma_{gg}/d\hat{t}$ by the sum of the singular terms in $\hat{t}$ and $\hat{u}$ (the last approximation in eq. (\ref{eq:sigmagg})).  Furthermore, $\hat{t}$ and $\hat{u}$ are symmetric under the integral so the integral can be done by keeping one of them and multiplying the result by 2.  The classical color field modification is characterized by setting $F_{mnj} = F_{mnj}^{cl}(x_1,x_2,Q^2=-\hat{t})$ in eq. (\ref{eq:mnj}).  The cross section is calculated by
\begin{eqnarray}
\sigma_{mnj} = &&\int_{\frac{2\hat{t}_0}{s}}^{1} dx_1 \int_{\frac{2\hat{t}_0}{x_1s}}^{1}dx_2 \int_{-\hat{s}+\hat{t}_0}^{-\hat{t}_0}d\hat{t}\, g(x_1,\mu_g^2)\, g(x_2,\mu_g^2) \nonumber \\ 
&& \times F_{mnj}^{cl}(x_1,x_2,Q^2=-\hat{t}) \frac{9\pi\alpha_s^2(\mu_g^2)}{\hat{s}^2}\left( -\frac{\hat{u}\hat{s}}{\hat{t}^2}\right)
 \label{eq:mnj2}
\end{eqnarray}
We use the running coupling 
\begin{equation}
\alpha_s(Q^2) = \frac{4\pi}{(11-8/3)\ln (Q^2/\Lambda_{QCD}^2)}
\end{equation}
where $\Lambda_{QCD} = 217$ MeV.  The hard cut-off scale is set to be $\hat{t}_0=1$ GeV$^2$.  We use the gluon distribution in \cite{Block2008}.  For the function $I(Q^2,x)$, we use $c=4.84$ as in the IIM model and set $\gamma = 0.9$.  So far, the minijet cross section includes only the $gg \rightarrow gg$ contribution that is adequate for very high energy.  But for smaller energy, one also needs to consider the contribution from qaurks scattering.  Empirically, the ratio between the total minijet cross section and the contribution from $gg \rightarrow gg$ can be approximated by 
\begin{equation}
\frac{\sigma_{\mbox{tot mnj}}}{\sigma_{gg\rightarrow gg}} = 1+ \frac{4}{{s}^{0.14}}.
\end{equation}
This parameterization is motivated by the result in \cite{Sarcevic1989}.  
For the soft component of eq. (\ref{eq:softpp}) and (\ref{eq:softpbp}), we use $\sigma_{soft} = 38.5$ mb and $a = 1.5$ GeV.  

\begin{figure}[h]
\centering
\includegraphics[width=1\linewidth]{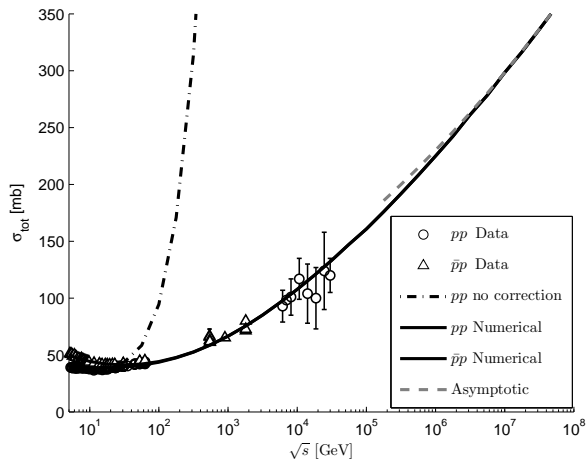}
\caption{\label{fig:4} Comparison between $pp$ and $\bar{p} p$ data with the modified minijet model.  The dash-dot line is calculated using the conventional minijet.  The solid lines are  $\sigma_{pp}$ and $\sigma_{\pbar p}$ of the present model.  The cross section of $\pbar p $ is higher than that of $pp$ at low energy $\sqrt{s}=5$ to 20 GeV.  The gray dashed line is the approximated asymptotic behavior of the total cross section.  The data is from \cite{PDG2010} and \cite{Antchev2011}.}
\end{figure}

We compare the total cross section to the $pp$ and $\bar{p} p$ data for energy $5$ GeV $ \le \sqrt{s} \le 30$ TeV.  The results of $pp$ and $\pbar p$ total cross sections from the present model are shown in Fig. \ref{fig:4} as the black lines with flat energy dependence and an initial decrease from $\sqrt{s}=5$ to $20$ GeV, respectively.  The conventional minijet cross section is also shown.  The figure shows that the presence of the classical color field suppresses the minijet cross section significantly.  

Note that, instead of optimizing the quality of the fit, our calculation is intended to show that, with the consideration of the effect of the classical color field from the high energy protons, the rise of the cross section can be satisfactorily described by the present model with a reasonable choice of parameters.  

\subparagraph*{\label{sec:4b}Asymptotic behavior of the $\sigma_{mnj}$:}
Beside the ability to fit the data, the present model exhibits a Froissart bound satisfying asymptotic behavior.  Consider the integral of eq. (\ref{eq:mnj2}) in the small $x_1$ and $x_2$ region where $Q_s^2(x)$ becomes large.  The dilogarithm $Li_2(-z)$ in eq. (\ref{eq:I}) is proportional to $\ln^2 (z)$ for large $|z|$.  So that the power law behavior of $Q_s^2(x)$ leads to a logarithmic behavior of $I(Q^2,x) \propto \ln(x)$ for small $x$.  Thus, $F_{mnj}^{cl}$, for small $x$, is 
\begin{equation}
F_{mnj}^{cl} \approx \frac{4 \chi^2 c^2}{\lambda^2\ln^2 \left(-\hat{t} R_p^2\right)\ln^2 x_1}\frac{4 \chi^2 c^2}{\lambda^2\ln^2 \left(-\hat{t} R_p^2\right)\ln^2 x_2}
\end{equation}
One can approximate the $\hat{t}$ integral by keeping only the most singular term ${1}/{\hat{t}^2}$.  The modified minijet cross section is given by
$\sigma^{asym}_{mnj} \equiv C H(s)$
where 
\begin{equation*}
C= \frac{144\pi \chi^4 c^4}{\lambda^4}\int_{-\infty}^{-\hat{t}_0} [C_2(-\hat{t})]^2\frac{\alpha_s^2(|\hat{t}|)}{\hat{t}^2 \ln^4 (|\hat{t}| R_p^2)} = 0.489 \mbox{mb}
\end{equation*}
with $C_2(Q^2)$ being the coefficient of $\ln^2 (1/x)$ of $g^{FB}(x,Q^2)$ in eq. (\ref{eq:gFB}) and
\begin{eqnarray*}
H(s) =  \int_{\frac{2\hat{t}_0}{s}}^1 dx_1\int_{\frac{2\hat{t}_0}{x_1s}}^1dx_2 \frac{\ln^2 \frac{1}{x_1} \ln^2 \frac{1}{x_2}}{x_1 x_2 \ln^2x_1 \ln^2x_2 } 
\xrightarrow{s\rightarrow \infty} \frac{\ln^2 s}{2}.
\end{eqnarray*}
Since our asymptotic approximation is done to the leading order in $\ln s$, it is correct up to a scale $s_0$.  In Fig. \ref{fig:4}, the asymptotic behavior of the total cross section, $\sigma_{tot}^{asym}(s) = \sigma_{soft} + (C/2) \ln^2 (s/s_0)$ with $s_0=0.7$ GeV$^2$ is plotted to show that the numerical calculation of the total cross section approaches to the asymptotic form.  Namely, $\sigma_{tot} \xrightarrow{s\rightarrow \infty} \sigma_{tot}^{asym} \propto \ln^2 s$.  Therefore, the present model gives a Froissart bound saturating cross section.  A similar asymptotic behavior were reported recently in \cite{Ishida2009,Block2011c}.

\subparagraph*{\label{sec:4c}Suppressed in the small x region:} 
At the tree level, the minijet is motivated by the exchange of gluon with $Q^2$.  This exchange gluon can be considered as a probe of the protons.  When the scattering involves collinear gluons at small $x$s, many partons with a higher $x$ value become a part of the source and the corresponding classical field increases.  The exchange gluon has to interact with the strong classical field before reaching the collinear gluon.  Throughout the process, the propagator acquires a suppressive factor which depends on $x$.  

\section{\label{sec:5}Conclusions}
The minijet model was originally introduced based on a striking resemblance between the rise of the total cross section with energy and the rise of jet (dominated by minijet) cross section. In this work, we show that within the framework of QCD,  the interaction between the gluon and the classical color field can play an important role in understanding this rise.

Among different parameterizations of the gluon distribution function, we choose the one that is consistent to the Froissart bound satisfying $F_2$ derived in \cite{Block2008}.  We first show that, even with this distribution, the conventional minijet cross section still violates the Froissart bound.  

We then apply the formalism of the gluon-classical color field interaction within the Gaussian approximation of CGC.  In this formalism, to leading order in the coupling constant and the strength of the sources, we found that the classical color field introduces a modification factor $F_{mnj}^{cl}$ in the integrand of the minijet cross section that leads to a suppression of the rate of the rise of the total cross section.  This allows a good description of the rise of the cross section comparing to the data as show in Figure \ref{fig:4}.  In addition, the total cross section has a $\ln^2 s$ asymptotic behavior which satisfies the Froissart bound.  
In this picture, the rise of the minijet cross section is still driven by the growth of gluon density in small $x$, but it is suppressed by the quantum-classical interaction from the dense medium.  

Although the complete solution of the classical color field due to two colliding protons has not yet found, our approximated solution to the leading order in the source, $\rho$ and coupling $g$ with Gaussian average illustrates a significant effect due to the quantum-classical interactions.  

In the present model, the collinear factorization is assumed.  In general, including the interaction with the classical field from the approaching protons would break the collinear factorization.  However, in \cite{Cheung2011a}, we found that the Feynman diagrams consisting of interactions between the collinear gluon and the classical field vanish.  The only correction to the $gg\rightarrow gg$ amplitude is of the form of eq. (\ref{eq:Fcl}) which is factorisable.  This justifies the factorization assumption.  Nevertheless, this factorization property is not guaranteed in the higher order calculation.  

\bibliography{./../paper_1/prd_2011_oct}

\end{document}